\documentclass[conference]{IEEEtran}
\IEEEoverridecommandlockouts
\usepackage{cite}
\usepackage{amsmath,amssymb,amsfonts}
\usepackage{algorithmic}
\usepackage{graphicx}
\usepackage{textcomp}
\usepackage{xcolor}
\usepackage{multirow}
\usepackage[super]{nth}

\makeatletter
\newcommand*{\rom}[1]{\expandafter\@slowromancap\romannumeral #1@}
\makeatother

\def\BibTeX{{\rm B\kern-.05em{\sc i\kern-.025em b}\kern-.08em
    T\kern-.1667em\lower.7ex\hbox{E}\kern-.125emX}}
\begin{document}

\title{Impact of Nap on Performance in Different Working Memory Tasks Using EEG
\footnote{{\thanks{This work was supported by the Institute of Information \& Communications Technology Planning \& Evaluation (IITP) grant, funded by the Korea government (MSIT) (No. 2019-0-00079, Artificial Intelligence Graduate School Program (Korea University)) and was partly supported by the Institute of Information \& communications Technology Planning \& Evaluation (IITP) grant, funded by the Korea government (MSIT) (No. 2021-0-02068, Artificial Intelligence Innovation Hub).}
}}
}

\makeatletter
\newcommand{\linebreakand}{%
  \end{@IEEEauthorhalign}
  \hfill\mbox{}\par
  \mbox{}\hfill\begin{@IEEEauthorhalign}
}
\makeatother

\author{\IEEEauthorblockN{Gi-Hwan Shin}
\IEEEauthorblockA{\textit{Dept. of Brain and Cognitive Engineering} \\
\textit{Korea University} \\
Seoul, Republic of Korea \\
gh\_shin@korea.ac.kr} 

\and

\IEEEauthorblockN{Young-Seok Kweon}
\IEEEauthorblockA{\textit{Dept. of Brain and Cognitive Engineering} \\
\textit{Korea University}\\
Seoul, Republic of Korea \\
youngseokkweon@korea.ac.kr}

\linebreakand 

\IEEEauthorblockN{Heon-Gyu Kwak}
\IEEEauthorblockA{\textit{Dept. of Artificial Intelligence} \\
\textit{Korea University} \\
Seoul, Republic of Korea \\
hg\_kwak@korea.ac.kr} 

\and

\IEEEauthorblockN{Ha-Na Jo}
\IEEEauthorblockA{\textit{Dept. of Artificial Intelligence} \\
\textit{Korea University} \\
Seoul, Republic of Korea \\
hn\_jo@korea.ac.kr} 

\and

\IEEEauthorblockN{Seong-Whan Lee}
\IEEEauthorblockA{\textit{Dept. of Artificial Intelligence} \\
\textit{Korea University} \\
Seoul, Republic of Korea \\
sw\_lee@korea.ac.kr} 
}

\maketitle

\begin{abstract} 
Electroencephalography (EEG) has been widely used to study the relationship between naps and working memory, yet the effects of naps on distinct working memory tasks remain unclear. Here, participants performed word-pair and visuospatial working memory tasks pre- and post-nap sessions. We found marked differences in accuracy and reaction time between tasks performed pre- and post-nap. In order to identify the impact of naps on performance in each working memory task, we employed clustering to classify participants as high- or low-performers. Analysis of sleep architecture revealed significant variations in sleep onset latency and rapid eye movement (REM) proportion. In addition, the two groups exhibited prominent differences, especially in the delta power of the Non-REM 3 stage linked to memory. Our results emphasize the interplay between nap-related neural activity and working memory, underlining specific EEG markers associated with cognitive performance.
\end{abstract}

\begin{small}
\textbf{\textit{Keywords--working memory, nap, electroencephalogram;}}\\
\end{small}

\section{INTRODUCTION} 
Sleep, a complex neurophysiological process integral to cognitive functioning, is divided into rapid-eye movement (REM) and non-REM (NREM) sleep, with the NREM sleep further subdivided into NREM 1 (N1), NREM 2 (N2), and NREM 3 (N3) stages \cite{eban2018neuronal, brodbeck2012eeg}. The N3 stage is particularly crucial for memory consolidation \cite{zhang2018electrophysiological}. Additionally, the concept of sleep architecture, which includes factors like total sleep time (TST), sleep onset latency (SOL), wake after sleep onset (WASO), sleep efficiency (SE), and the relative proportions of N1, N2, N3, and REM stages, is intricately linked to memory processes \cite{alger2010delayed}. Yet, the differential impacts of these sleep characteristics on various memory tasks remain to be fully explored.

Many studies have focused on working memory, a crucial element for complex cognitive tasks such as reasoning and learning \cite{shin2021predicting,small2001circuit}. Using electroencephalography (EEG), known for its high temporal resolution, has been instrumental in revealing the intricate working memory processes \cite{ref10,ref9}. The advantage of EEG in capturing rapid neural activity related to cognition is particularly evident in spectrogram analysis \cite{shin2023changes}. This method offers comprehensive insights into oscillatory brain activity across various frequency bands, including delta (1-4 Hz), theta (4-8 Hz), alpha (8-13 Hz), beta (13-25 Hz), and gamma (25-50 Hz), which enhances our understanding of the neural oscillations associated with different components of working memory \cite{ferrarelli2019increase,genzel2014light}. However, the specific effects of different sleep stages on distinct working memory tasks remain less understood, marking an area ripe for further investigation.

In this study, we aimed to identify the effects of a nap on word-pair and visuospatial working memory tasks. We hypothesized that the impact of the nap could vary across these tasks, particularly expecting pronounced effects across various stages of sleep. To investigate this, we grouped participants using a clustering method based on their performance in working memory tasks, considering factors like accuracy and reaction time. Our analysis began with a comparison of sleep architecture between the groups to discern any differences in sleep patterns and their potential effects on memory tasks. Subsequently, we conducted spectrogram-based EEG analyses to examine neural activity during the naps, aiming to correlate these patterns with working memory performance.

\begin{figure*}[t!]
\centering
\scriptsize
\includegraphics[width=\textwidth]{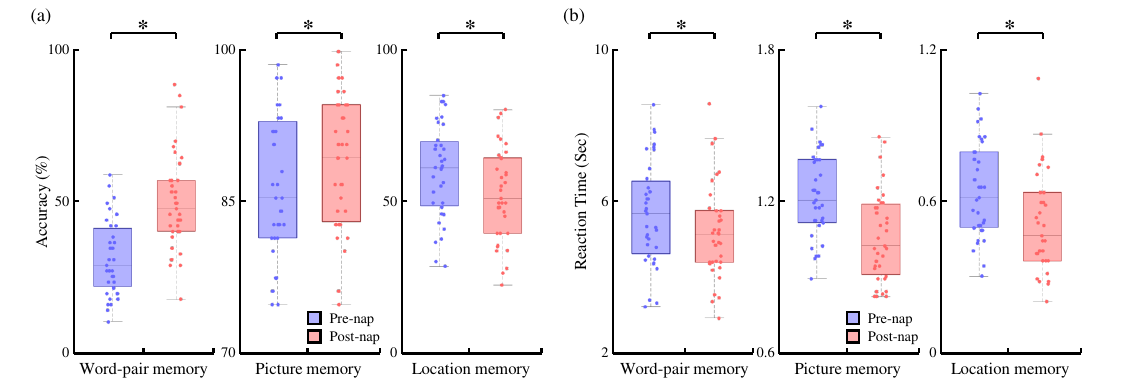}
\caption{Performance of distinct working memory tasks in pre- and post-nap sessions. For (a), accuracy rate is shown, and for (b), reaction time is displayed. Black asterisks indicate statistical significance at \textit{p}$<$0.05.}
\end{figure*}

\section{METHODS} 
\subsection{Participants and Experimental Setup} 
Thirty-five participants (14 females, age: 25.4 ± 2.3 years) were recruited after providing written informed consent. None had neurological or psychiatric disorders. Following the procedure outlined by Shin \textit{et al.} \cite{shin2020assessment}, participants underwent three sessions: pre-nap, a 90-minute nap, and post-nap. During the pre- and post-nap sessions, they completed two working memory tasks: a word-pairs task \cite{leminen2017enhanced} and a visuospatial task involving picture and location memory \cite{ladenbauer2016brain}. The tasks were developed using Psychtoolbox (http://psychtoolbox.org) \cite{ref7}. The study received approval from the Institutional Review Board at Korea University (KUIRB-2021-0155-03).

\subsection{Data Recording and Preprocessing} 
Data were recorded from 60 EEG and 4 electrooculography (EOG) channels using 64 electrodes and the BrainAmp system (Brain Products GmbH, Germany). Electrodes were positioned according to the 10-20 system for EEG and placed near the eyes for EOG. The data were sampled at 1,000 Hz with FCz and Fpz as the reference and ground. All electrode impedances were maintained below 20 k$\Omega$.

EEG signals were preprocessed using EEGLAB \cite{delorme2004eeglab} and BCILAB toolboxes \cite{kothe2013bcilab} for MATLAB 2018b. Signals were down-sampled to 250 Hz and band-pass filtered between 1 and 50 Hz \cite{ref8}. Eye movement artifacts were corrected using independent component analysis with the EOG signals. A Laplacian spatial filter was applied to improve the signal-to-noise ratio \cite{ref4}.

Sleep experts labeled the data into five sleep stages (wake, N1, N2, N3, and REM) using 30-second epochs, following American Academy of Sleep Medicine standards \cite{silber2007visual}.

\subsection{Data Analysis} 
\subsubsection{Cognitive performance analysis} 
To identify the impact of nap on various working memory, participants undertook tasks pre- and post-nap sessions, with their performance measured in terms of accuracy and reaction time. Using a clustering method, we divided participants into high-performers and low-performers for each working memory task, based on these performance metrics. This approach aimed to highlight variations in nap effects linked to proficiency in each task.

\begin{table*}[t!]
\caption{Sleep Architecture of High- and Low-Performers Based on Performance in Distinct Working Memory Tasks (Mean $\pm$ SD).}
\centering
\renewcommand{\arraystretch}{1.7}
\begin{tabular*}{\textwidth}{@{\extracolsep{\fill}\quad}lccccccccc}
\hline
& \multicolumn{3}{c}{Word-pair   memory}      & \multicolumn{3}{c}{Picture   memory} & \multicolumn{3}{c}{Location   memory}        \\ \cline{2-10}
                                                                               & HP           & LP           & \textit{p}-value             & HP            & LP           & \textit{p}-value     & HP           & LP           & \textit{p}-value              \\ \hline
TST (min)                                                                      & 70.29 $\pm$ 0.75 & 79.64 $\pm$ 1.16 & 0.098         & 77.56 $\pm$ 0.76  & 72.81 $\pm$ 0.71 & 0.461 & 70.10 $\pm$ 0.90 & 79.27 $\pm$ 0.85 & 0.101          \\ 
SOL (min)                                                                      & 5.60 $\pm$ 0.14  & 3.21 $\pm$ 0.25  & \textbf{0.035} & 5.06 $\pm$ 0.30   & 4.50 $\pm$ 0.14  & 0.671   & 5.15 $\pm$ 0.17  & 3.97 $\pm$ 0.21  & 0.303          \\
WASO (min)                                                                     & 14.62 $\pm$ 0.69  & 7.64 $\pm$ 1.00 & 0.166         & 7.89 $\pm$ 0.60   & 13.19 $\pm$ 0.63 & 0.351 & 15.25 $\pm$ 0.79 & 7.27 $\pm$ 0.75  & 0.108          \\
SE (\%)                                                                        & 78.10 $\pm$ 0.83 & 88.49 $\pm$ 1.29 & 0.098         & 86.17 $\pm$ 0.85  & 80.90 $\pm$ 0.79 & 0.461 & 77.89 $\pm$ 1.00 & 88.07 $\pm$ 0.94 & 0.101          \\
N1 (\% of TST)                                                                 & 29.95 $\pm$ 0.86 & 25.08 $\pm$ 1.50 & 0.468         & 23.13 $\pm$ 1.13  & 29.69 $\pm$ 0.82 & 0.383  & 30.85 $\pm$ 1.04 & 24.20 $\pm$ 1.10 & 0.316          \\
N2 (\% of TST)                                                                 & 50.61 $\pm$ 0.79 & 58.38 $\pm$ 1.18 & 0.182         & 56.02 $\pm$ 1.40  & 52.93 $\pm$ 0.69 & 0.640  & 50.29 $\pm$ 0.91 & 58.30 $\pm$ 0.91 & 0.164          \\
N3 (\% of TST)                                                                 & 15.92 $\pm$ 0.94 & 10.36 $\pm$ 1.00 & 0.370         & 13.23 $\pm$ 1.59  & 13.86 $\pm$ 0.73 & 0.928 & 16.70 $\pm$ 0.99 & 9.70 $\pm$ 0.95  & 0.253          \\
R (\% of TST)                                                                  & 3.51 $\pm$ 0.32  & 6.18 $\pm$ 0.66  & 0.326         & 7.63 $\pm$ 0.85   & 3.52 $\pm$ 0.29  & 0.175 & 2.16 $\pm$ 0.25  & 7.80 $\pm$ 0.65  & \textbf{0.031} \\ \hline
\end{tabular*}
\end{table*}

\begin{figure*}[t!]
\centering
\scriptsize
\includegraphics[width=\textwidth]{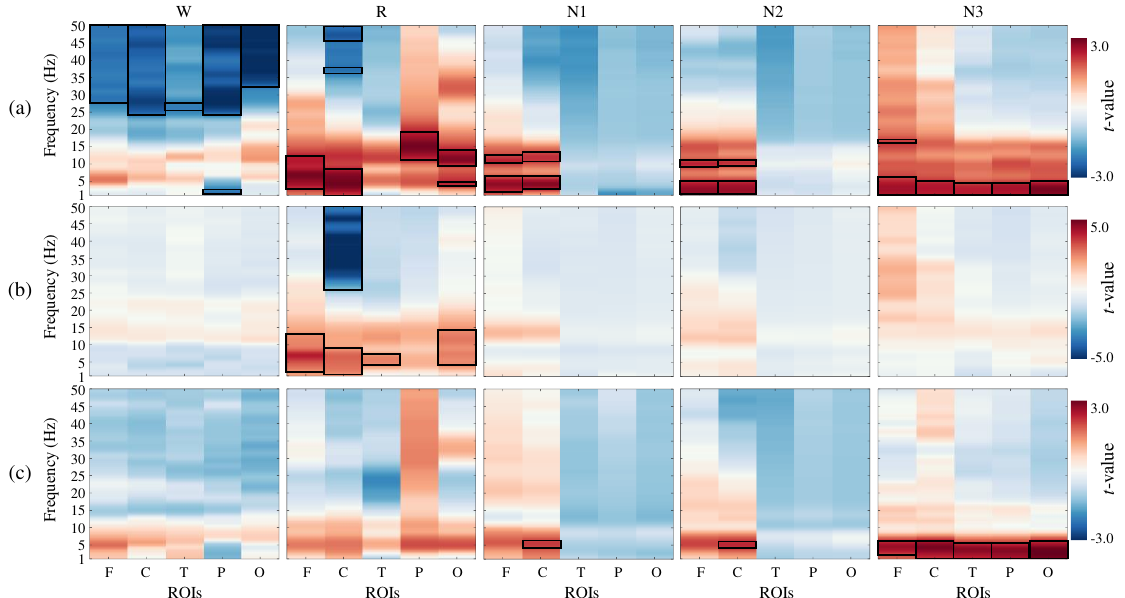}
\caption{Statistical maps illustrating the relative increase and decrease in spectrogram between high- and low-performers during nap session. For (a) word-pair memory, (b) picture memory, and (c) location memory, the channels corresponding to the Regions of Interest (ROIs) were averaged for each of the five sleep stages, highlighting the frequency differences within the range of 1-50 Hz. The black-bordered regions indicate significant frequencies and ROIs (\textit{p}$<$0.05).}
\end{figure*}

\subsubsection{EEG analysis} 
EEG data were analyzed via spectrograms to capture the neural oscillatory patterns prevalent across different sleep stages in nap sessions. The analysis encompassed a frequency range of 1 to 50 Hz. Time-domain signals were converted into frequency-domain representations, employing a 99 \% window overlap to ensure high temporal resolution \cite{ref6}. This methodology was instrumental in identifying dominant frequencies and elucidating their distribution patterns throughout the various sleep stages. Furthermore, we categorized the EEG channels into five key regions: frontal, central, temporal, parietal, and occipital, to investigate a region-specific analysis of oscillatory patterns \cite{ref5}.

\subsection{Statistical Analysis} 
To compare the changes in working memory accuracy and reaction time for pre-nap and post-nap sessions, a paired \textit{t}-test was applied. In addition, a two-sample \textit{t}-test was used to assess the nap's impact on high- and low-performer groups by separately analyzing their sleep architecture and EEG data. For all these statistical evaluations, a significance threshold was set at \textit{p}$<$0.05.

\section{RESULTS}
\subsection{Influence of Nap on Working Memory Performance}
We conducted a comparative analysis of performance in distinct working memory tasks during pre- and post-nap sessions, as depicted in Fig. 1. In accuracy, there was a significant improvement post-nap in both word-pair and picture memory tasks, while the location memory task showed a notable decrease. In reaction time, a significant reduction was observed across all working memory tasks in the post-nap session.

\subsection{Variations in Sleep Architecture Across Memory Tasks}
To elucidate the impact of napping on distinct working memory tasks, we conducted a statistical analysis of sleep architecture between high- and low-performers, as detailed in Table \rom{1}. In the word-pair memory task, high-performers statistically exhibited a longer SOL compared to low-performers. Conversely, in the location memory task, low-performers showed a significantly higher proportion of REM sleep. For the picture memory task, no significant statistical differences in sleep architecture were observed between the two groups.

\subsection{EEG Analysis During Nap and Memory Task Performance}
In our EEG analysis, as depicted in Fig. 2, we observed varied patterns across memory tasks and performer groups during the nap. In the word-pair memory task, low-performers exhibited higher delta power in the parietal region and increased gamma power in all brain regions during the wake stage, with higher gamma power in the central region during the REM stage. In contrast, high-performers demonstrated elevated powers in various frequency bands, excluding the gamma band, in different brain regions during the REM stage, and higher powers in delta, theta, and alpha bands during the N1, N2, and N3 stages. For the picture memory task, the REM stage showed notable differences, with low-performers exhibiting increased gamma power in the central region, whereas high-performers had higher powers in other low-frequency bands. In the location memory task, high-performers displayed elevated theta power in the central region during the N1 and N2 stages and higher delta and theta powers across all brain regions in the N3 stage.

\section{DISCUSSION}
In the current study, our research investigated the impact of napping on distinct working memory tasks using EEG. We observed significant differences in accuracy and reaction time between pre- and post-nap sessions in participants. Based on these results, we categorized participants into high- and low-performer groups and identified significant variations in SOL and the proportion of REM. Furthermore, an EEG comparison of the two groups across different working memory tasks revealed distinct variations in neural patterns across a variety of frequency bands and brain regions. These results highlight the complex interplay between sleep, neural activity, and working memory performance, indicating that nap can differentially affect cognitive processing in diverse brain regions.

This study conducted statistical comparisons to unravel the impacts of nap on different cognitive performances. The observed improvement in accuracy for word-pair and picture memory tasks in post-nap sessions suggests the benefits of nap in memory consolidation \cite{walker2004sleep}. However, the decrease in location memory observed in post-nap sessions suggests a contrasting effect, indicative of memory loss or interference \cite{alvarez2022napping}. Additionally, we observed a significant reduction in reaction time across all working memory tasks in post-nap sessions, highlighting the restorative effects of nap on overall alertness \cite{milner2009benefits,rasch2013sleep}. These findings emphasize that naps significantly influence the performance of various working memory tasks.

Our analysis reveals group differences in sleep architecture and its impact on various working memory tasks. High performers in the word-pair memory task showed a longer SOL, suggesting a complex interplay between sleep dynamics and verbal memory processing, though this doesn't necessarily imply more effective memory consolidation \cite{zhang2022role}. Conversely, in the location memory task, the higher proportion of REM sleep in low-performers questions its role in location memory consolidation, indicating that increased REM sleep doesn't uniformly enhance performance \cite{tilley1978rem}. In the picture memory task, the absence of significant sleep differences between high- and low-performers points to the influence of factors beyond sleep architecture, like individual cognitive strategies or inherent memory abilities \cite{diekelmann2010memory}. Collectively, these findings highlight the intricate relationship between sleep architecture and working memory and suggest that understanding the impact of sleep stages on memory could aid in addressing cognitive decline \cite{ref3}.

The EEG analysis in our study uncovered distinct neural activity patterns associated with napping's effects on various working memory tasks, revealing a complex interplay among sleep stages, brain activity, and cognitive performance. In the word-pair memory task, low-performers showed higher delta power in the parietal region during the wake stage and increased gamma power in the central region during REM sleep, indicating intense neural processing. High-performers, however, demonstrated increased delta, theta, and alpha powers in deeper sleep stages, particularly in the delta band of the N3 stage, which is closely associated with memory processes. This pattern suggests a critical role of N3 in memory consolidation \cite{holz2012eeg}. In the picture memory task, differences in the REM stage were noted, with low-performers showing increased gamma power in the central region, while high-performers exhibited higher powers in low-frequency bands \cite{vijayan2017frontal}. The location memory task showed that high-performers had elevated theta power in the central region during N1 and N2 stages and higher delta and theta powers in N3 stage across all brain regions \cite{shin2020assessment}. These findings highlight napping's impact on different working memory.

The study's limitation is the non-randomized order of working memory tasks, potentially affecting the results. Future research plans to implement a randomized sequence and apply machine learning to better understand how sleep stages influence cognitive performance, aiming to tailor cognitive enhancement to individuals \cite{ref1,ref2}. Insights from such studies have the potential to inform strategies not only for cognitive enhancement but also for addressing memory deficits associated with dementia.

\section{CONCLUSION}
Our research provides insights into how nap affects performance across various working memory tasks. The observed differences across performance groups highlight the potential role of nap in cognitive function. The EEG data, particularly from the NREM stage, reveal distinct neural patterns associated with various working memory tasks. These EEG characteristics suggest the neural mechanisms involved in memory processing during sleep. These findings lay a foundation for future studies, enriching our understanding of the interplay between sleep and cognitive functions.

\bibliographystyle{IEEEtran}
\bibliography{REFERENCE}

\end{document}